\documentclass{aastex62}

\newcommand{\heii}{\mbox{He\, \textsc{ii}}}
\newcommand{\niii}{\mbox{N\, \textsc{iii}}}

\newcommand{\vesini} {\mbox{\textrm{$v_{\textrm e}\sin{i}$}}}

\newcommand{\kms} {\mbox{\textrm{km$\;$s$^{-1}$}}}

\newcommand{\teff}{\ensuremath{T_{\rm eff}}}
\newcommand{\logg}{\ensuremath{\log{g}}}

\received{}%January 1, 2018}
\revised{\today}%January 7, 2018}
\accepted{}
\submitjournal{ApJL}

\shorttitle{The fastest rotator}
\shortauthors{Li et al.}
\usepackage{graphicx}
\begin{document}

\title{LAMOST J040643.69+542347.8: the fastest Rotator in the Galaxy}

\correspondingauthor{Guang-Wei Li}
\email{lgw@bao.ac.cn}

\author[0000-0001-7515-6307]{Guang-Wei Li}
\affiliation{Key laboratory of Space Astronomy and Technology, National Astronomical Observatories, Chinese Academy of Sciences, 
Beijing 100101, China}



\begin{abstract}
%Massive stars influence the star formation and chemical evolution of galaxies, especially in the early Universe. 
Rotation and binary interaction play important roles in understanding the nature 
of massive stars ($\gtrsim 8 M_\odot$). Some interesting transients, such as 
the long-duration gamma-ray bursts, are thought to be originated from 
fast-rotating massive stars. Because the strong stellar wind can effectively spin 
down a metal-rich massive star with fast rotation, it is very hard to find single 
massive stars rotating critically in the Galaxy. In the present work reported is the 
discovery of the fastest rotator in the Galaxy, LAMOST J040643.69+542347.8, 
with a projected rotational velocity $\vesini \sim540$ km\,s$^{-1}$,  which is 
$\sim 100$ km\,s$^{-1}$ faster than that of the previous record holder 
HD~191423. The star has a spectral type of O6.5 Vnnn(f)p. Its He I 
$\lambda$4471 absorption line is blueshifted and asymmetric, while its He II 
$\lambda$4686 and H$\alpha$ have central absorption reversals in their 
emissions. It is also a runaway star, which implies an origin in a close binary 
interaction.  Compared to VFTS 285 and VFTS 102 (their $\vesini \sim 610$ 
km\,s$^{-1}$) in the Large Magellanic Cloud, LAMOST J040643.69+542347.8 
has its own peculiar spectral characteristics and earlier spectral type. Moreover, 
LAMOST J040643.69+542347.8 is bright (B $\sim13.9$ mag) enough to allow 
future high-resolution spectroscopic follow-ups.

\end{abstract}


%% From the front matter, I move on to the body of the paper.
%% I recommend that authors also use the natbib \citep

\section{Introduction} \label{sec:intro}
The fast rotation of a massive star can induce the mixing that can transpose 
the fresh fuel H into the core, while bringing the material produced by the CNO 
cycle in the stellar core onto the surface, which enriches the surface 
abundance, prolongs the stellar life, and increases the stellar luminosity 
\citep{mae87,mey00,bro11,lan12}. Rotational mixing can produce, if 
strong enough, chemically homogeneous evolution (CHE). However, the 
strong stellar wind in a metal rich massive star can effectively brake the rotation 
and therefore the mixing process (e.g. see Fig. 15 in \citet{mae12}). As 
a result, it is hard to find a fast rotator born as a single star in high metallicity 
regions.
\par
Fortunately, most of massive stars are in binary systems \citep{mas09,chi12}. 
\citet{san12} proposed that $71\%$ of O-type stars would exchange materials 
with their companions during their whole lifetime. Binary interaction, through 
tides and/or by inducing mass and angular momentum accretion onto the 
secondary can produce fast-rotating stars \citep{pac81,lan12,de13}. In fact, 
high-velocity components in the distributions of $\vesini$ for single O- and 
B-type stars in 30 Dor are found by \citet{ram13} and \citet{duf13}, respectively, 
which are speculated to be the products of binary interactions \citep{de13}. 
Moreover, \citet{li20} found that most of the O stars with extreme N enrichment 
(i.e. ON stars) are runaways with very fast rotation speeds, which implies their 
origins in binary interactions. 

\par
So far, the fastest rotators are VFTS 285 and VFTS 102  with $\vesini \sim 
610$ \kms~\citep{ram13,duf11,wal12}, which are located in the Large 
Magellanic Cloud (LMC). For comparison, the previous fastest rotator in the 
Galaxy is HD~191423, with a projected rotation speed of only 435 
\kms~\citep{how01}. In fact, it is noteworthy that all three of these stars are all 
runaways, which imply their binary origin.

\par

In this Letter, the fastest rotator LAMOST J040643.69+542347.8 in the Galaxy 
is presented, with the projected rotational velocity $\vesini \sim 540$ 
km\,s$^{-1}$, which is also a runaway. Further study on LAMOST 
J040643.69+542347.8, as well as VFTS 285 and VFTS 102, in theory and 
observation can help us understand the origin and nature of these extreme 
rotators, and also the natures of the progenitors of gravitational-wave (GW) 
events \citep{de09,de16} and long-duration gamma-ray bursts (LGRBs) 
\citep{yoo05,woo06,can17}.

\section{Data and Analysis} \label{sec:data}
The star LAMOST J040643.69+542347.8 was a serendipitous discovery  
when we searched for Oe-type stars \citep{li18} in LAMOST low-resolution 
spectra (LRSs) \citep{wang96,su04,cui12,luo12,zhao12}. Its LAMOST spectrum 
is shown in Fig. \ref{fig:lamost}. The  resolving power of LRSs is about 1,800,  
while the spectral dispersion around He II $\lambda$4542 is about 
$0.583 ~\rm{\AA}/$pixel.

\subsection{Spectral Type}

\par
\begin{figure*}
%\begin{center}
\includegraphics[angle=0, scale=0.9]{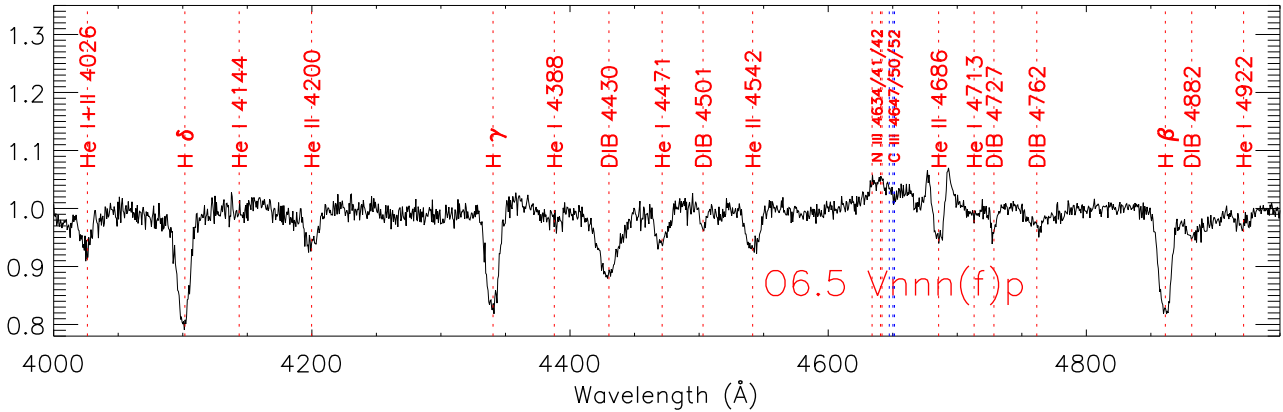}
\caption{The LAMOST spectrum of LAMOST J040643.69+542347.8. Diffuse interstellar bands (DIBs) are also indicated.
\label{fig:lamost}}
%\end{center}
\end{figure*}

The criteria given by \citet{sot11} are used to assign its spectral type. 
From Fig. \ref{fig:lamost} we can see that its He~I $\lambda$4471 is slightly 
weaker than He~II $\lambda$4542 (also see Panel B of Fig. \ref{fig:rot_profile}), 
so the star LAMOST J040643.69+542347.8 is classified as spectral type O6.5. 
Its unusually broad lines indicate that it is a very fast rotator, while its weak 
\niii\ $\lambda\lambda$4634--4640--4642 emissions indicate that it is also an 
O(f) star. Moreover, there is an absorption in He II $\lambda$4686 emission, 
which is the classic characteristic of the Onfp category 
\citep{wal73,wal10,sot11}, and the emission obviously weakens the 
absorption, so the \heii~$\lambda$4686 absorption should be stronger than it 
looks. \citet{wal01} showed that \heii~$\lambda$4686 is sensitive to surface
gravity. The strong absorption of  \heii ~$\lambda$4686 implies that the $\logg$ 
of some parts of this star's surface is very high. Thus, it should be a dwarf. As a 
result, it is assigned a spectral type of O6.5 Vnnn(f)p. 
\par

\subsection{Distance and Peculiar Velocity}
 \label{sec:dis}
The method given by \citet{li20} is used to deduce the distance and peculiar 
velocity from data of Gaia DR2 \citep{gaia2}. The distances to the Sun and the 
Galactic center are $8901.8^{+5147.1}_{-1565.2}$ pc and $16386.9^{+5020.5}
_{-1498.5}$ pc, respectively, which imply that this star is located in the Outer 
Arm of the Galaxy. Its peculiar velocity $(V_{\rm r}, V_{\rm c}, V_{\rm z}) = 
(97.4^{+48.7}_{-16.0}, -48.6^{+15.3}_{-14.6}, 44.2^{+22.7}_{-6.9})$ \kms, and 
$V = \sqrt{V_{\rm r}^2+ V_{\rm c}^2+ V_{\rm z}^2 } = 118.3^{+47.3}
_{-12.7}$~\kms, where, $V_{\rm r}, V_{\rm c}$ and $V_{\rm z}$ are the peculiar 
Galactocentric radial, circular, and vertical velocities, respectively. The velocity 
is greater than 28 \kms, which is the threshold for runaway stars 
given by \cite{tet11}. Therefore, this star probably was originated from a binary 
system, then kicked off by a nonspherically symmetric explosion of its companion.

\subsection{Radial Velocity, Projected Rotational Velocity, and Atmospheric  Parameters}
The method given by \citet{li20} is used to determine radial and projected 
rotational velocities ($v_{\textrm{R}}$ and $v_{\rm e} \sin i$) from the \heii\ 
$\lambda$4542, because this line should be formed in the deep stellar 
photosphere and marginally influenced by the stellar wind \citep{de04}. 
The resulting radial velocity $v_{\textrm{R}} = -91 \pm 14$ \kms.
\par
To obtain its $\vesini$, the model 
\begin{equation}
\label{equ:line}
F(\lambda,\vesini)= a \times S(\lambda)\otimes P(\lambda)\otimes G(\lambda,\vesini) + b
\end{equation}
is used to fit its \heii\ $\lambda$4542 profile.
In Equation \ref{equ:line}, $\otimes$ is the convolution operation, $S(\lambda)$ 
is the spectrally intrinsic profile without rotation, $P(\lambda)$ is the instrument 
profile, $G(\lambda,\vesini)$ is the rotational-broadening profile \citep{gray05}
for a given $\vesini$, $a$ is the intensity of the line, and $b$ is the continuum. 
\par
The instrument profile $P(\lambda)$ is from the arc lamp spectrum that is 
used for wavelength calibration for the spectrum of this star. There are only 
a few emission lines in the arc lamp, so I use the linear interpolation of the 
profiles of two neighbor emission lines at \heii\ $\lambda$4542 as 
$P(\lambda)$.
\par
The wings of H$\beta$ and H$\gamma$ lines in Figure \ref{fig:lamost} are very 
narrow and steep, which implies a low gravity, while the unusually broad lines 
imply it is a nearly edge-on extreme rotator. In fact, the centrifugal force and 
radiative pressure can effectively reduce its total gravity near the equator 
\citep{mae00}. 
\par
\citet{liang19} presented that the Galaxy has a radial metallicity gradient in the 
thin disk, and at the place where this star is located, the metallicity is similar to 
that of the LMC. However, the fast rotation can induce the mixing, which implies 
that the surficial metallicity might be enriched. Thus, the \heii\ ~$\lambda$4542 
profiles in the Potsdam Wolf-Rayet (PoWR) grids \citep{hai19} of LMC and solar 
metallicity with 31 kK $\leqslant \teff \leqslant$ 42 kK and 3.4 dex 
$\leqslant \logg \leqslant$ 4.0 dex are used as the $P(\lambda)$s in Equation 
\ref{equ:line} to calculate $\vesini$. All the resulting $\vesini$ are 
greater than 500~\kms.
\par
\begin{figure*}
%\begin{center}
\includegraphics{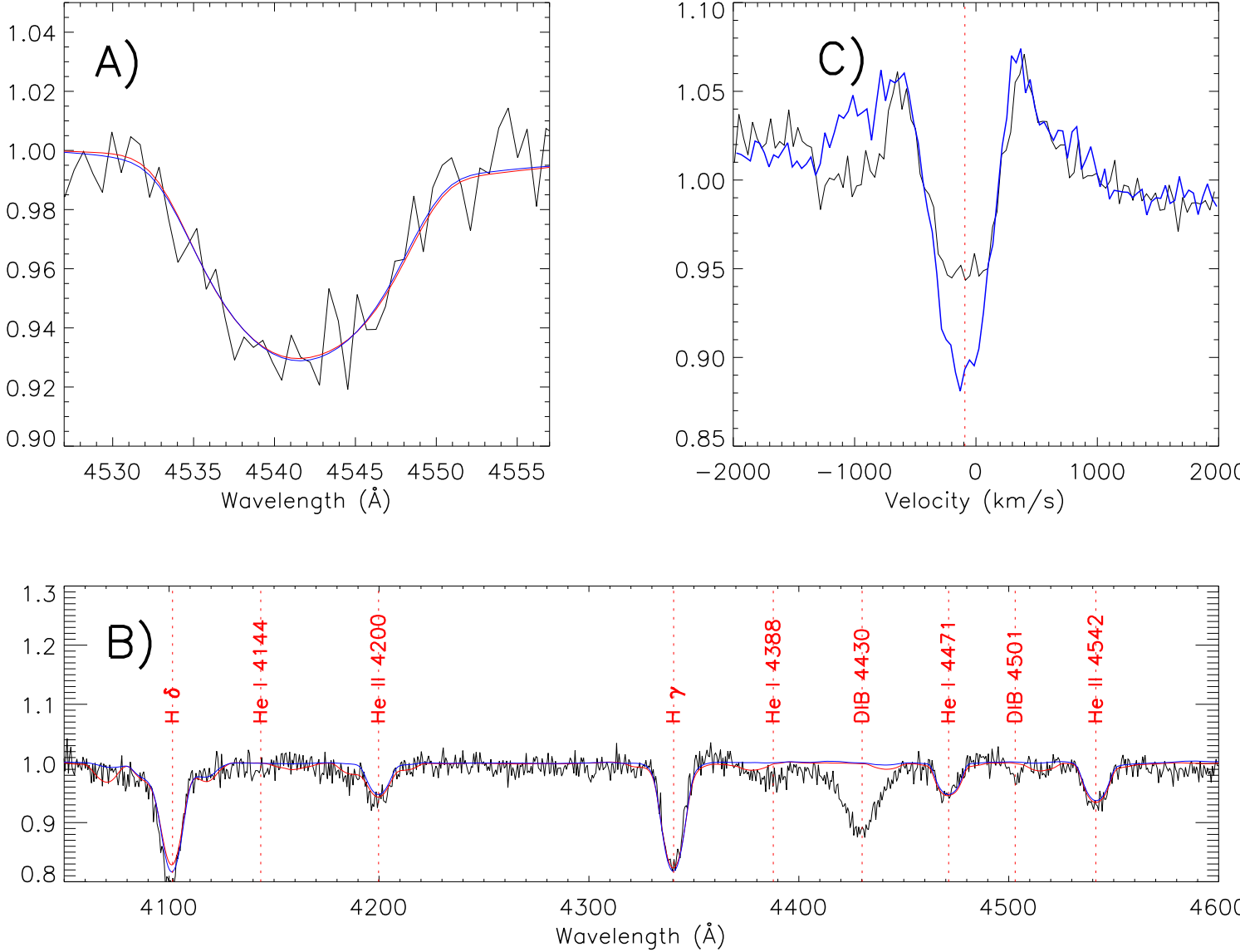}%{cmp_tpl_spec.eps}
\caption{Panel A: The black line is the He II $\lambda$4542 profile in the 
LAMOST spectrum, while the fitted profiles calculated from two best PoWR 
spectra with LMC and solar metallicity  are shown in blue and red, respectively. 
Panel B: The two best-fitted spectra with LMC and solar metallicity are shown in 
blue and red, respectively, while the LAMOST spectrum is shown in black. 
Panel C: The profiles of \heii~$\lambda$4686 and H$\alpha$ in the LAMOST 
spectrum are shown in black and blue, respectively.
\label{fig:rot_profile}}
%\end{center}
\end{figure*}

Each PoWR spectrum is convolved with the $\vesini$ calculated from it and the 
instrument profile, then compared with the LAMOST spectrum. The best 
spectrum with LMC metallicity has $\teff = 35$ kK and $\logg = 3.6$ dex, while 
the best spectrum with solar metallicity has $\teff = 34$ kK and $\logg = 3.6$ 
dex. These two best spectra are shown in blue and red, respectively, in Panel B 
of Fig. \ref{fig:rot_profile}, while $\vesini$ calculated from their 
\heii~$\lambda$4542 are $543 \pm 29$ ~\kms and $538 \pm 29$~\kms, 
respectively, and their corresponding rotational profiles are also shown in blue 
and red in Panel A of Fig. \ref{fig:rot_profile}. Because of gravity darking 
\citep{von24}, $\vesini$ may be underestimated by even several tens of 
percents \citep{tow04}. 
\par

\subsection{Variablity}
\begin{figure*}
%\begin{center}
\includegraphics[angle=90, scale=0.5]{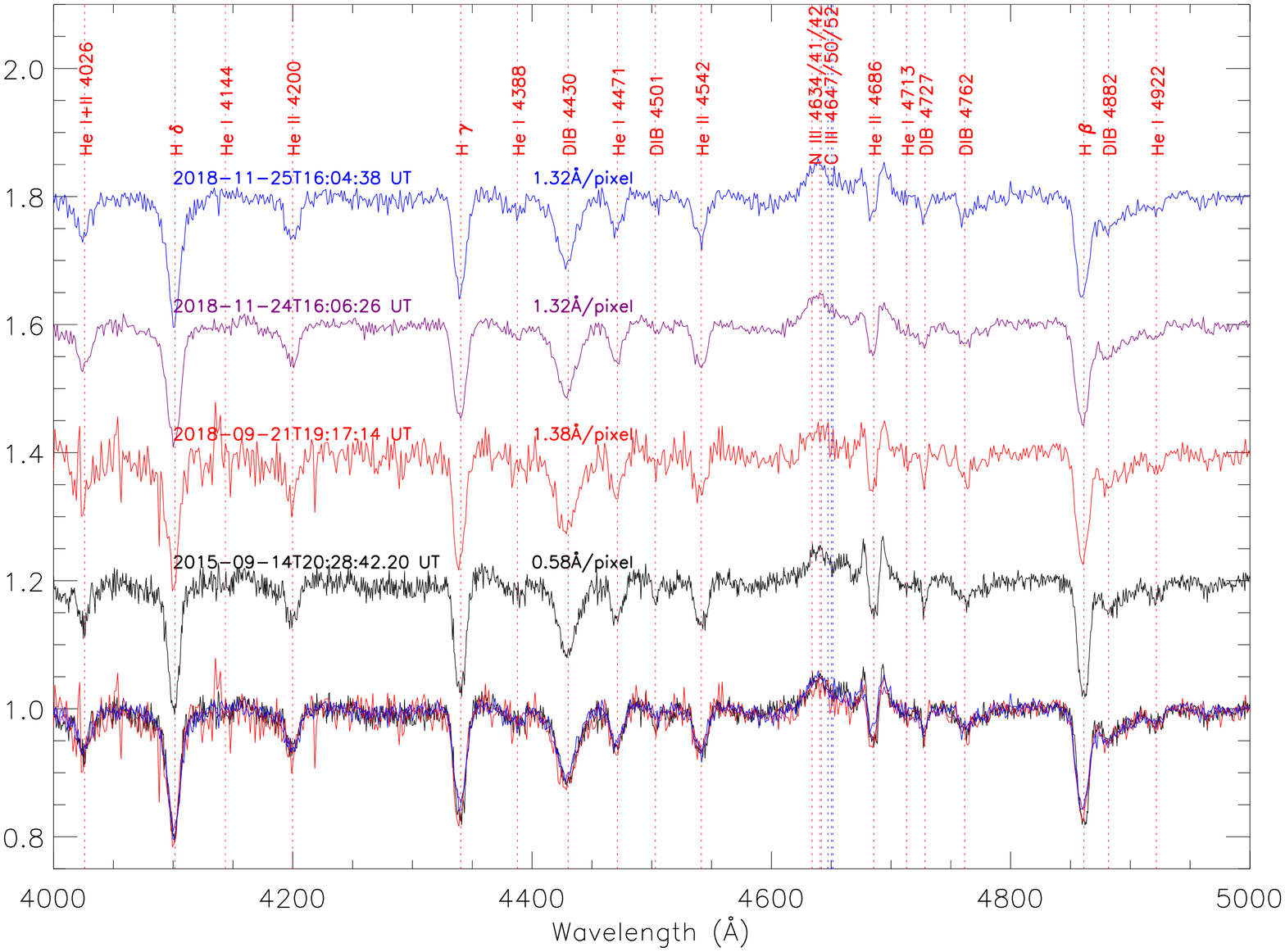}
\caption{The upper three spectra were obtained with the 2.16m telescope in 
Xinglong station, China, while the black spectrum was obtained with LAMOST. 
The observation time and dispersion are indicated above each spectrum. All 
spectra are overplotted together at the bottom by the same colors as the above.
\label{fig:spec}}
%\end{center}
\end{figure*}

Another three LRSs were obtained from 2.16m Telescope at 
Xinglong Observatory in China with two instruments: Beijing-Faint Object 
Spectrograph and Camera (BFOSC) and the spectrograph made by 
Optomechanics Research Inc. for low-resolution spectroscopy (OMR). The 
spectra are shown in Figure \ref{fig:spec}. He~II $\lambda$4686 shows obvious 
variability between four spectra, but the variability of radial velocity remains 
uncertain because of the low resolutions and signal-to-noise of the spectra. 
\par
The light curves of the star LAMOST J040643.69+542347.8 are found in the 
data archives of the Zwicky Transient Facility (ZTF) \citep{mas19} and the Wide 
Angle Search for Planets (WASP)\citep{but10}, but there are no photometric 
variations larger than their photometric errors.
 
\subsection{Luminosity}
I obtained the theoretical $B-V = -0.276$ and $M_V = -5.752$ mag from the 
flux-calibrated PoWR spectrum of LMC metallicity with $\teff = 35$ kK and 
$\logg = 3.6$ dex, where the B and V filter profiles are from \citet{man15}. The 
flux-calibrated PoWR spectrum of solar metallicity with $\teff = 34$ kK and 
$\logg = 3.6$ dex also has a similar $B-V$ and $M_V$ as the above.
\par
In APASS DR9 \citep{hen15}, the star LAMOST J040643.69+542347.8 has 
$B = 13.069\pm0.108$ mag and $V = 13.912\pm0.102$ mag, respectively. 
Thus, if the standard extinction coefficient $R=3.1$ is adopted, then by using 
the Markov Chain Monte Carlo method, its observational $M_{\rm V} = 
-5.43^{+0.78}_{-1.05}$ mag. 
\par

\section{Discussion}
In Panel C of Fig. \ref{fig:rot_profile}, He II $\lambda$4686 and H$\alpha$ have
similar profiles: the narrow absorptions in broad emissions, which implies that 
they have a similar origin. In fact, He II $\lambda$4686 and H$\alpha$ are 
all sensitive to gravity \citep{wal01} in O-type stars: they are in emissions in the 
photosphere with a low gravity, but in absorptions in the photosphere with a 
high one. LAMOST J040643.69+542347.8 is an extreme rotator, so its  
equatorial radius is larger than the polar radius, which results in higher 
gravities at poles while a lower one around the equator. Therefore, I speculate 
that their broad emissions may form from low latitude, because of low gravity 
and fast rotation there, and even in the stellar wind, while narrow absorptions 
may form near the poles because of high gravity and low rotation there. 
\par
He I $\lambda$4471 is asymmetric and blue-shifted and keeps constant for 
more than three years, as shown in Fig. \ref{fig:spec}. \citet{gag19} present that 
for an extreme rotator, the stellar wind around the equator is stronger 
than other places because of the bistability jump. Thus, I speculate that He I 
$\lambda$4471 may be affected by the wind originating around the 
equator. If it is true for LAMOST J040643.69+542347.8, then this star has 
much stronger losses of mass and angular momentum around the equator, 
which would result in a rapid spin down of the rotation in the future. 
\par
For comparison, the VFTS 285 \citep{wal12} and VFTS 102 \citep{duf11} are 
the two fastest rotators in the LMC, with $\vesini \sim 610$ 
\kms~\citep{ram13}. Their spectral types are O7.5 Vnnn \citep{wal12} and 
O9 Vnnne \citep{duf11}, respectively, with normal He I $\lambda$4471 and 
He II $\lambda$4686 absorption profiles. However, the slow and dense wind 
around the equator is still found in VFTS 285, while the double-peaked 
H$\alpha$ emission in the spectrum of VFTS 102  indicates that the material is 
being ejected from the equator and forms a circumstellar disk \citep{she20}. 
These phenomena show the angular momentum losses are braking down the 
rotations of these two stars.
\par
Moreover, \citet{song16} presented that binary interaction can trigger CHE. In 
fact, the overenrichments of N and He and depletions of C and O on have 
been found on the surface of HD 191423 \citep{vil02}, which is the second 
fastest rotator in the Galaxy now, with $\vesini \sim 435$~\kms~\citep{how01}, 
and also a runaway star. Therefore, it would be interesting to obtain 
higher-resolution spectra to determine the surface composition of this star. 

\section{Conclusion} 
In this Letter, I report the fastest rotator LAMOST J040643.69+542347.8 in 
the Galaxy found with LAMOST. Its properties  are listed in 
Table \ref{tab:property}.
 \par
LAMOST J040643.69+542347.8 is a runaway, which implies it originates 
from binary interaction. I cannot find the obvious variabilities of its radial 
velocity and photometries, which implies that it may have no (at least 
sufficiently massive) companion. Its blueshifted and asymmetric 
He I $\lambda$4471 and absorption reversals in He II $\lambda$4686 and 
H$\alpha$ emissions are unique characteristics in the extreme rotator category. 
\par
LAMOST J040643.69+542347.8 is bright enough to allow future high-resolution 
spectroscopic follow-up, so it is a good laboratory to examine theories related to 
fast rotation, e.g. CHE, bistability jump of the wind around the equator, 
progenitors of LGRBs and GW events.

\startlongtable
\begin{deluxetable}{ll}
\tablecaption{Observational and Theoretical Properties \label{tab:property}}
\tablehead{
\colhead{Property} & \colhead{Estimate}
}
\startdata
\hline
R. A. ($\alpha$) & 61.682058$^{\circ}$ \\
Decl. ($\delta$) & $54.396626^{\circ}$ \\
Gaia $\varpi$ & $0.1125\pm0.0286$ mas \\
Gaia $\mu_{\alpha}$ & $-0.816\pm0.048$ mas yr$^{-1}$ \\
Gaia $\mu_{\delta}$	& $2.002\pm0.036$	mas yr$^{-1}$ \\\hline
Gaia G  & $12.6678\pm0.0003$ mag \\
Gaia BP & $13.1996\pm0.0016$ mag\\
Gaia RP & $11.9591\pm0.0010$ mag\\
B & $13.912\pm0.102$ mag \\
V  & $13.069\pm0.108$ mag \\	
SDSS u & $14.760\pm0.004$ mag\\
SDSS g &	$14.922\pm0.009$ mag\\
SDSS r &	$12.917\pm0.002$ mag\\
SDSS i & $12.492\pm0.001$ mag\\
SDSS z & $12.831\pm0.005$ mag\\	
2MASS J & $11.063\pm0.021$ mag\\
2MASS H & $10.864\pm0.020$ mag\\
2MASS K$_s$& $10.720\pm0.020$ mag\\\hline
%WISE W1& $10.613\pm0.023$ mag\\
%WISE W2 & $10.592\pm0.021$ mag\\
%WISE W3 & $10.705\pm0.089$ mag\\\hline
%WISE W4	& $8.951
% g=13.43, r=12.71, i=12.35) (ra: 61.682058, dec: 54.396626)
Spectral Type & O6.5 Vnnn(f)p \\
$\vesini$ & $540\pm29$~\kms \\
$v_{\rm R}$ & $91\pm 14$~\kms \\
$\teff$ & $35 \pm 1$ kK \\
$\logg$ & $3.6 \pm 0.2$ dex \\
$d$ & $8901.8^{+5147.1}_{-1565.2}$ pc \\
peculiar velocity &$ 118.3^{+47.3}_{-12.7}$~\kms\\
$M_V$ & $-5.43^{+0.78}_{-1.05}$ mag \\
\enddata

\end{deluxetable}

\acknowledgments
I thank the anonymous referee, who provided detailed and valuable feedback that substantially improved the Letter.
\par
I thank Rodolfo Hector Barba Suarez for valuable advices. I thank Nolan Walborn (deceased), Sergio Sim{\'o}n-D{\'{\i}}az, Chris Evans, Ian Howarth, and Nidia Morrell for their discussions.
\par
This research is supported by the National Natural Science Foundation of China (NSFC; grant No. 11673036).  
\par
Guoshoujing Telescope (the Large Sky Area Multi-Object Fiber Spectroscopic Telescope LAMOST) is a National Major Scientific Project built by the Chinese Academy of Sciences. Funding for the project has been provided by the National Development and Reform Commission. LAMOST is operated and managed by the National Astronomical Observatories, Chinese Academy of Sciences.
\par
I acknowledge the support of the staff of the Xinglong 2.16m telescope. This work was partially supported by the Open Project Program of the Key Laboratory of Optical Astronomy, National Astronomical Observatories, Chinese Academy of Sciences.


\vspace{5mm}
\facilities{Guoshoujing Telescope(LAMOST), GAIA, ZTF, 2.16m telescope at Xinglong Station}


\software{astropy \citep{2013A&A...558A..33A},  
          emcee \citep{for13}
          }

\end{document}